\documentclass[apj]{emulateapj}
\usepackage{amsmath}
\usepackage{graphicx}
\usepackage{amssymb}
\usepackage{times}
\newcommand{\na}{\rm New Astron.}
\newcommand{\pasa}{\rm PASA}

\begin{document}

\title{On the origin of fast radio bursts (FRBs)}

\author{Eli Waxman\altaffilmark{1}\altaffilmark{2}}
\altaffiltext{1}{Dept. of Particle Phys. \& Astrophys., Weizmann Institute of Science, Rehovot 76100, Israel}
\altaffiltext{2}{I would like to thank the Institute for Advanced Study, Princeton, for its hospitality during a period in which a significant part of this work has been carried out.}
\begin{abstract}

We derive stringent constraints on the persistent source associated with FRB 121102: Size $0.3<R_{17.5}=(R/10^{17.5}~{\rm cm})<3$, age $<10^{2.5}$~yr, energy $E\approx10^{49}(\varepsilon_e/0.2~{\rm GeV})^3$~erg, characteristic electron energy $0.1\le\varepsilon_e/{\rm 1~GeV}\le0.5$; The radiating plasma is confined by a cold plasma of mass $M_c<10^{-1.5}R_{17.5}^4M_\odot$. These properties are inconsistent with typical "magnetar wind nebulae" model predictions.
The fact that $\varepsilon_e\sim m_p c^2$ suggests that the hot plasma was created by the ejection of a mildly relativistic, $M\approx E/c^2\approx 10^{-5}M_\odot$ shell, which propagated into an extended ambient medium or collided with a pre-ejected shell.

Independent of the persistent source model, we suggest a physical mechanism for the generation of FRBs: Ejection from an underlying compact object, $R_s=10^{6}R_{s,6}$~cm, of highly relativistic shells, with energy $E_s=10^{41}E_{41}$~erg and Lorentz factor $\gamma_s = 10^3 E_{41}^{1/8}R^{-3/8}_{s,6}$, into a surrounding $e-p$ plasma with density $n\sim10^{-1}{\rm cm^{-3}}$ (consistent with that inferred for the persistent source). For $E_s$ similar to observed FRB energies, plasma conditions appropriate for strong synchrotron maser emission at $\nu_{\rm coh.}\approx 0.5 E_{41}^{1/4}R^{-3/4}_{s,6}{\rm GHz}$ are formed. A significant fraction of the deposited energy is converted to an FRB with duration $R_s/c$, accompanied by $\sim10$~MeV gamma-rays carrying less energy than the FRB.

The inferred energy and mass associated with the source suggest some type of a "weak stellar explosion", where a neutron star is formed with relatively low mass and energy ejection. However, the current upper limit on $R$ does not allow one to rule out $M_c\sim1M_\odot$, or the ejection of larger mass well before the ejection of the confining shell.

\end{abstract}

\maketitle

\section{Introduction}
\label{sec:intro}

Fast radio bursts (FRBs) are short, $\Delta t\lesssim10^{-3}$~s, bright, $0.1-10$~Jy, bursts of radio waves in the GHz range \citep{2007Sci...318..777L,2012MNRAS.425L..71K,2013Sci...341...53T,2014ApJ...790..101S,2015ApJ...799L...5R,2016PASA...33...45P,Scholz16Repeating,2016MNRAS.460L..30C}. Their sources and the mechanism responsible for their production are yet unknown \citep[see e.g.][for review]{Katz16Rev}.

The sub-arc-second localization \citep{Chatterjee17FRB_loc} of the repeating FRB 121102 \citep{Spitler16Repeat,Scholz16Repeating} lead to the identification of a dwarf galaxy at a redshift of $z=0.19$ \citep{2017ApJ...834L...7T} and of a persistent radio source \citep{Chatterjee17FRB_loc}, both located in a direction consistent with that of the FRB.

In \S~\ref{sec:persistent} we show that stringent constraints on the properties of the persistent source are obtained under the assumption, that the persistent source and the FRB source are physically associated and reside at the dwarf galaxy. Our approach differs from that of earlier analyses, which attempt to interpret the emission in terms of an assumed model of the source \citep[e.g.][]{Kashiyama17,Metzger17Magnetar,Beloborodov17Magnetar,Dai17Magnetar_no_ejecta,Katz17dark,Piro17massive_ejecta,Katz17Lightning,Zhang17comb}, typically of emission from a "magnetar wind nebula" confined by a supernova ejecta. We derive constraints on the parameters of the plasma producing the persistent emission, which are independent of assumptions implied by adopting a specific underlying model for the source.

Our analysis goes beyond that of \citet{Beloborodov17Magnetar}, who derived the electron number density and the total energy, $E$, assuming specific values for the source size, $R=10^{17}$~cm, age, $t_{\rm persist}=30$~yr, and magnetic field to electron energy density, $\epsilon_B/\epsilon_e\approx 1$. We show that the values of $R$, $t_{\rm persist}$ and $\epsilon_B/\epsilon_e$ are constrained by the observed dispersion measure and by the lack of its variation, and derive the allowed range of parameters of both the radiating and the confining plasmas(under the assumption that the FRB source resides within the persistent source, the dispersion measure of the FRBs constrains the properties of the plasma producing the persistent emission).

The high brightness temperatures of FRBs suggest that they are produced by a coherent emission mechanism operating within an unstable plasma configuration. It should be noted, that the energy emitted in a single FRB event of the repeating FRB 121102, $\sim10^{39}$~erg, is $\sim10^{-10}$ of the energy stored in the plasma producing the persistent emission ($E\sim10^{49}$~erg as we show below), and that the average luminosity carried by FRBs is $\sim10^{-5}$ of the persistent luminosity \citep{Spitler16Repeat,Chatterjee17FRB_loc}. It is unlikely therefore that the global properties of the source could be inferred from the properties of the FRBs, or that a unique mechanism for coherent radio emission due to some unstable plasma configuration could be singled out. Nevertheless, we suggest in \S~\ref{sec:FRB} a plausible mechanism for coherent emission, consistent with the properties of the FRBs (a brief discussion of other proposed mechanisms is given in \S~\ref{summary:FRB}).

The mechanism is based on the modification of synchrotron emission in the presence of a plasma with a refractive index that deviates from unity, $n^2=1-(\nu_p/\nu)^2$, where $\nu_p$ is the plasma frequency (see \S~\ref{sec:maser}). The possibility of negative reabsorption (i.e. amplification) of synchrotron radiation by relativistic electrons in the presence of a cold (non-relativistic) plasma has been considered by various authors following \citet{McCray66} and \citet{Zheleznyakov67} \citep[see][for review]{Ginzburg89}. \citet{Sazonov70} has shown that negative reabsorption is possible in the absence of a cold plasma, due to the modification of $n$ by the relativistic electron plasma. He found that this is possible at $\nu\le\gamma_e\nu_p$, where $\gamma_e$ is the Lorentz factor of the electrons, provided $\gamma_e^2\ll \nu_p/\nu_B$, where $\nu_B$ is the electron gyration frequency. \citet{SW02} have shown that such amplification is possible also for $\gamma_e^2>\nu_p/\nu_B$, and that in general negative reabsorption may be obtained at $\nu\le\nu_{R^*}= \min[\gamma_e,(\nu_p/\nu_B)^{1/2}]\nu_p$ ($\nu_{R^*}$ is a generalization of the Razin-Tsytovich frequency to the regime $\gamma_e^2>\nu_p/\nu_B$).

We show in \S~\ref{sec:FRB} that deposition of energy characteristic of FRB events over a time scale characteristic of FRB events into an $e-p$ plasma with density $\sim10^{-1}{\rm cm^{-3}}$, leads to conditions appropriate for strong negative reabsorption of synchrotron emission at $\sim 1$~GHz, which may convert a significant fraction of the deposited energy into a synchrotron maser radio emission.

The model suggested for FRB generation is independent of the model for the persistent source. However, as we show below, the density inferred for the plasma producing the persistent emission is consistent with the density of the plasma required for generating the FRBs by our suggested mechanism.

Our conclusions and additional predictions are disucssed in \S~\ref{sec:discussion}.

\section{The persistent source}
\label{sec:persistent}

In \S~\ref{sec:persistent_size} we derive upper limits on the persistent source size based on its observed variability. In \S~\ref{sec:persistent_const} and \S~\ref{sec:persistent_prop} we derive the plasma properties based on the observed properties of the radio emission. The key observed properties of the source, used in our analysis, are listed below.

\begin{enumerate}
  \item The luminosity and angular distances to the source, assuming that it resides in the dwarf galaxy, are $d_L=970$~Mpc and $d_A=680$~Mpc respectively.
  \item Emission from the persistent source was observed over 150~d. The FRB source is active for $>4$~yr.
  \item The dispersion measure, $DM$, of all FRB events is consistent with $558\pm3{\rm pc/cm^3}$ \citep{Spitler16Repeat,Chatterjee17FRB_loc}.
  The contribution of the local environment of the source to the $DM$ is $\le200{\rm pc/cm^3}$ \citep{2017ApJ...834L...7T}.
  \item The VLBI angular size, 0.2 and 2~mas at 5 and 1.7~GHz respectively \citep{Marcote17size}, is consistent with broadening due to scattering, $\theta\propto\nu^{-11/5}$. Using the size at 5~GHz, the source size should satisfy $R\ll0.7$~pc$=2\times10^{18}$~cm.
  \item At 3~GHz, the source shows 10\% to $30$\% variability on $\sim10$~d time scale \citep{Chatterjee17FRB_loc}.
  \item The radio flux peaks at $\sim10$~GHz, with $\nu F_\nu\cong2\times10^{-17}{\rm erg/cm^2s}$, corresponding to $\nu L_\nu\cong2\times10^{39}{\rm erg/s}$ \citep{Chatterjee17FRB_loc}.
  \item The flux extends approximately like $\nu F_\nu\propto \nu^1$ down to $\sim1$~GHz \citep{Chatterjee17FRB_loc,Marcote17size}.
\end{enumerate}

\subsection{Variability constraints on the source size}
\label{sec:persistent_size}

If the variability of the persistent source was intrinsic to the source, it would imply a source size $R<10^{17}$~cm. However, the variability is consistent with being due to refractive scintillation, which would be obtained also for a larger source size. The angular size, $\theta_d$, variability time scale, $t_s$, and maximum RMS flux variation due to refractive scintillation of a source of a finite size $\theta_s\ll\theta_d$ are \citep[][sec 3.3]{Goodman97}
\begin{equation}\label{eq:scint_size}
    \theta_d=
    0.2 (\nu/5~{\rm GHz})^{11/5}(SM_{-3.5}/80)^{3/5}{\rm mas},
\end{equation}
\begin{equation}\label{eq:scint_time}
    t_s=22\frac{\theta}{0.2~{\rm mas}}\left(\frac{v}{50~{\rm km/s}}\right)^{-1}\frac{d}{\rm kpc}\,{\rm d},
\end{equation}
\begin{equation}\label{eq:scint_mp}
    m_{\rm sc.,peak}=0.13\left(\frac{\theta_s}{0.1~{\rm mas}}\right)^{-17/66}\left(\frac{SM_{-3.5}}{80}\right)^{-1/22},
\end{equation}
\begin{equation}\label{eq:scint_nup}
    \nu_{\rm peak}=3\left(\frac{\theta_s}{0.1~{\rm mas}}\right)^{-5/11}\left(\frac{SM_{-3.5}}{80}\right)^{3/11}.
\end{equation}
Here, the scattering measure is normalized as $SM=10^{-3.5}SM_{-3.5}{\rm kpc/m}^{20/3}$, $d$ is the scattering screen distance and $\nu_{\rm peak}$ is the frequency at which the maximum RMS fractional flux variation $m_{\rm sc.,peak}$ is obtained ($m_{\rm sc.}\propto\nu^{17/30}$ below $\nu_{\rm peak}$, and $m_{\rm sc.}\propto\nu^{-2}$ above; For given $SM$, $\nu_{\rm peak}$ increases as $\theta_s$ decreases, but $m_{\rm sc.}$ below the peak does not change).

From Eq.~(\ref{eq:scint_size}) we infer $SM_{-3.5}\approx80$, consistent with $SM$ values of lines of sight within the disk and in the direction of the FRB \citep{Marcote17size}. The observed variations imply $m_{\rm sc.}(3~{\rm GHz})\cong0.15$, which in turn imply, using eqs.~(\ref{eq:scint_mp}) and (\ref{eq:scint_nup}), $\theta_s<0.1$~mas and
\begin{equation}\label{eq:Rmax}
  R<10^{18}~{\rm cm}.
\end{equation}
This result suggests that the observed variability is not intrinsic and indeed due to scintillation. As noted above, the source size, that would be inferred under the assumption that the variability is predominantly intrinsic, is $R<10^{17}$~cm. Eq.~(\ref{eq:Rmax}) implies that for such a source size the RMS flux variations due to scintillation would be consistent (for the inferred $SM$) with the observed variations, which implies that the intrinsic variability does not contribute significantly.

\subsection{Constraints on the properties of the persistent source}
\label{sec:persistent_const}

Let us assume that the radiation is produced by synchrotron emission of a plasma sphere of radius $R$, electron density $n_e$, electron Lornetz factor $\gamma_e$, and magnetic field $B$. The observed properties set stringent constraints on these parameters.

We first demonstrated that the radiating electrons must be highly relativistic, and that the source does not expand relativistically.
\begin{enumerate}
    \item The limit on the source size, $R<10^{18}$~cm, and the fact that it is active for $>4$~yr implies that it cannot be expanding relativistically, $R/t_{\rm persist}<10^{10}{\rm cm/s}$.
    \item The fact that no self-absorption is observed at low frequency (in the persistent source spectrum) implies
    \begin{equation}\label{eq:Tb}
        4\pi R^2\frac{2\nu^3}{c^2}\gamma_e m_e c^2>\nu L_\nu,
    \end{equation}
    from which we obtain (using $\nu L_\nu=3\times10^{38}{\rm erg/s}$ at 1.7~GHz)
    \begin{equation}\label{eq:gTb}
      \gamma_e\ge10^{1.5}R_{17.5}^{-2}.
    \end{equation}
    Here, $R=10^{17.5}R_{17.5}$~cm.
\end{enumerate}

Using the peak flux and frequency of the radio emission, we may now determine two source parameters ($B$ and $n_e$) as function of the two others ($R$, $\gamma_e$). Identifying the peak at 10~GHz with the synchrotron frequency,
\begin{equation}\label{eq:nus_eq}
    \nu_s=\gamma_e^2\frac{eB}{2\pi m_e c}
\end{equation}
and
\begin{equation}\label{eq:L_eq}
    \nu L_\nu=\frac{4\pi}{3}n_eR^3\frac{4}{3}\sigma_T c\gamma_e^2 U_B,
\end{equation}
we find
\begin{equation}\label{eq:nus}
    B=10^{-1.5}\gamma_{e,2.5}^{-2}{\rm G}
\end{equation}
and
\begin{equation}\label{eq:L}
    n_e=0.1R_{17.5}^{-3}\gamma_{e,2.5}^{2}{\rm cm}^{-3}.
\end{equation}
Here, $\gamma_e=10^{2.5}\gamma_{e,2.5}$.

The spectrum at $\nu<10$~GHz is consistent with emission from the electrons for which $\nu_s=10$~GHz, with no significant cooling (the fact that the observed spectrum is somewhat softer than the $\nu L_\nu\propto\nu^{4/3}$ behavior expected at $\nu\ll\nu_s$ is consistent with the interpretation that the observed frequencies are close to $\nu_s$). This sets a lower limit to the cooling time, $E_e/L>t_{\rm persist}$ where $t_{\rm persist}$ is the age of the source, $>4$~yr. This in turn sets a lower limit to $\gamma_e$,
\begin{equation}\label{eq:ts}
    \gamma_{e,2.5}>0.8 t_{9}^{1/3},
\end{equation}
where $t_{\rm persist}=10^9t_9$~s. This limit is independent of the non-absorption limit, and is consistent with it.

The plasma is relativistic, but does not expand relativistically. It should therefore be confined by some (cold) shell with density $\rho=n_c m_p$, such that the velocity of the shock driven by the hot plasma into the confining plasma, $\sqrt{U/n_c m_p}$ where $U\cong\max\{U_B,U_e\}$ is the plasma energy density, is smaller than $R/t_{\rm persist}$. This yields
\begin{equation}\label{eq:n}
    n_c>n_{\rm min}=10^2\max\left[2.8\gamma_{e,2.5}^{-4},1.8\gamma_{e,2.5}^{3}R_{17.5}^{-3}\right]R_{17.5}^{-2}t_{9}^2\,\rm cm^{-3}.
\end{equation}

The dispersion measure induced by the shocked part of the denser outer shell is $\sqrt{U/n_cm_p}n_ct_{\rm persist}\propto n_c^{1/2}$ (to set conservative constraints, we consider the contribution to the $DM$ of only the shocked part of the confining shell, which is heated and ionized by the shock; the un-shocked plasma may be neutral). The minimum dispersion is obtained for the minimum density, for which the shock travels a distance $R$. Thus
\begin{eqnarray}\label{eq:DM}
    DM&\ge&\xi n_{\rm min}R\cr
    &=&20\xi\max\left[1.5\gamma_{e,2.5}^{-4},\gamma_{e,2.5}^{3}R_{17.5}^{-3}\right] R_{17.5}^{-1}t_{9}^2\,\rm \frac{pc}{cm^{3}}.
\end{eqnarray}
Similarly, the change in $DM$ over the $\delta t=4$~yr period, $\sqrt{U/n_cm_p}n_c\delta t\propto n_c^{1/2}$, satisfies
\begin{eqnarray}\label{eq:dDM}
    \delta DM&\ge&\xi n_{\rm min}R\delta t/t_{\rm persist}\cr &=&2\xi\max\left[1.5\gamma_{e,2.5}^{-4},\gamma_{e,2.5}^{3}R_{17.5}^{-3}\right]R_{17.5}^{-1}t_{9}\,\rm \frac{pc}{cm^{3}}.
\end{eqnarray}
Here, $\xi$ is a dimensionless parameter of order unity, the value of which depends on the exact geometry and dynamics (for a shock wave propagating into a uniform medium, for example, $\xi=1/3$).

Requiring the contribution of the shocked part of the confining cold shell to the $DM$ and its variation, $\delta DM$, not to exceed $100~{\rm pc/cm^3}$ and $3~{\rm pc/cm^3}$ respectively, eqs.~({\ref{eq:DM}) and~(\ref{eq:dDM}) set upper limits (and a new lower limit) on $\gamma_e$, which may be combined with the earlier lower limits to give
\begin{eqnarray}\label{eq:gDM}
    &\max&\left[0.1R_{17.5}^{-2},0.7\xi^{1/4}R_{17.5}^{-1/4}t_9^{1/2},0.8t_{9}^{1/3}\right]\cr
    &<& \gamma_{e,2.5}< 1.7 \xi^{-1/3}t_{9}^{-2/3} R_{17.5}^{4/3},
\end{eqnarray}
and
\begin{eqnarray}\label{eq:gdDM}
    &\max&\left[0.1R_{17.5}^{-2},1\xi^{1/4}R_{17.5}^{-1/4}t_9^{1/4},0.8t_{9}^{1/3}\right]\cr
    &<& \gamma_{e,2.5} < 1.1 \xi^{-1/3}t_{9}^{-1/3} R_{17.5}^{4/3}.
\end{eqnarray}
We do not combine eqs.~(\ref{eq:gDM}) and~(\ref{eq:gdDM}) into a single inequality in order to demonstrate that the limits obtained from the total $DM$ and from its variation are similar.

A self consistent solution exists only for
\begin{equation}\label{eq:t1}
    t_{9}<2\xi^{-1/2}\min\left[\xi^{1/6} R_{17.5}^{4/3},R_{17.5}^{19/14},35R_{17.5}^{5}\right],
\end{equation}
and
\begin{equation}\label{eq:t2}
    t_{9}<\min\left[1.6\xi^{-1/2} R_{17.5}^{2},1.2\xi^{-1}R_{17.5}^{19/7},1.3\times10^3\xi^{-1}R_{17.5}^{10}\right].
\end{equation}

\subsection{The properties of the persistent source}
\label{sec:persistent_prop}

To summarize, we may draw the following conclusions regarding the the properties of the persistent source.
\begin{enumerate}
    \item Eqs.~(\ref{eq:t1}) and~(\ref{eq:t2}) imply $t_{\rm persist}<10^{2.5}$~yr.
    \item Eq.~(\ref{eq:t2}) implies that $R>10^{17}$~cm, as otherwise the maximal age $t_{\rm persist}$ is smaller than $4$~yr.
    \item Eqs.~(\ref{eq:gDM}) and~(\ref{eq:gdDM}) imply that $0.6\lesssim\gamma_{e,2.5}\lesssim3$.
    \item The energy contained in the electrons and the ratio of magnetic to electron energy are
    \begin{equation}\label{eq:Ee}
        E_e=10^{48.5}\gamma_{e,2.5}^3{\rm erg},
    \end{equation}
    \begin{equation}\label{eq:UBUe}
        \frac{E_B}{E_e}=1\gamma_{e,2.5}^{-7}R_{17.5}^3.
    \end{equation}
    \item If the plasma is neutral due to the presence of protons, its mass is
    \begin{equation}\label{eq:Mv}
        M=10^{-5}\gamma_{e,2.5}^2M_\odot.
    \end{equation}
    The observations are therefore consistent with the emission of a $\sim10^{-5} M_\odot$ shell at mildly relativistic speed, with $E_k\sim Mc^2\sim10^{49}$~erg. The conversion of the kinetic energy to thermal energy by a collisionless shock driven by the collision with the denser surrounding medium would convert the kinetic energy to thermal energy, with characteristic energy per particle of $\sim m_p c^2$ and magnetic field not far below equipartition.
    \item An upper limit on the density of the confining plasma is set by the requirement that its contribution to $DM$ and $\delta DM$, $DM\approx\sqrt{Un_c/m_p}t_{\rm persist}$ and $\delta DM\approx\sqrt{Un_c/m_p}\delta t$, be smaller than $100{\rm pc/cm^3}$ and $3{\rm pc/cm^3}$ respectively. The resulting upper limits on the density and mass, $M_c\approx(4\pi/3)R^3n_cm_p$, of the confining plasma are approximately given by
    \begin{equation}\label{eq:n_c}
        n_c<10^{2.5}R_{17.5}{\rm cm}^{-3},
    \end{equation}
    \begin{equation}\label{eq:M_c}
        M_c<10^{-1.5}R_{17.5}^4M_\odot.
    \end{equation}
    Here too, to set conservative constraints we consider the contribution to the $DM$ of only the shocked part of the confining shell, which is heated and ionized by the shock; the un-shocked part may be neutral.
\end{enumerate}

\section{A model for the FRBs}
\label{sec:FRB}

We consider the ejection of a highly relativistic shell, with Lorentz factor $\gamma_s$ and kinetic energy $E_s=\gamma_s M_s c^2$, from a central compact object of size $R_s$ into the plasma producing the persistent emission. The expanding shell drives a forward shock into the $n_e\sim0.1{\rm cm^{-3}}$ plasma (see Eq.~(\ref{eq:L})), and the pressure behind the forward shock drives a reverse shock into the expanding shell. We show below that if the plasma producing the persistent emission is an $e-p$ plasma, with $n_p=n_e=n$, then the plasma conditions behind the reverse shock are appropriate for strong stimulated amplification of synchrotron emission.

We show below, that a significant fraction of the kinetic energy $E_s$ is converted to internal energy within the shocked shell plasma, and may therefore be radiated as an FRB, at the radius $r_d$ where the shell begins to decelerate. We therefore derive first, in \S~\ref{sec:dec}, the characteristic plasma parameters at $r_d$ and the frequency at which negative reabsorption is expected, and then discuss stimulated ("maser") emission in \S~\ref{sec:maser}.

The following point should be emphasized here. Since the ejected shells are required to be highly relativistic, $\gamma_s\gg1$, the ejection of shells which are not spherical, but rather conical sections of a sphere with opening angle $\theta_s\gg 1/\gamma_s$, would lead to similar observed burst properties (as long as our line of sight to the source falls within the cone): A conical shell with $\theta_s\gg 1/\gamma_s$ behaves as if it were part of a spherical shell (until significant deceleration sets in). Thus, $E_s$ should be regarded as the "spherically equivalent energy", while the true energy emitted is $\sim\theta_s^2 E_s$.

\subsection{Plasma conditions at the deceleration radius}
\label{sec:dec}

At small radii, the proper density of the expanding shell, $n_s$, is large and the reverse shock is weak. As the shell expands, its density decreases, and the reverse shock becomes stronger. The Lorentz factor, $\gamma$, and pressure, $p$, of the plasma lying between the reverse and forward shocks are roughly uniform (with a contact (density) discontinuity separating the plasma shocked by the reverse and forward shocks). Since the reverse shock decelerates the expanding shell, $\gamma<\gamma_s$.

Let us assume that, as we show later to be valid, the reverse shock never becomes highly relativistic. In this case, $\gamma\approx\gamma_s$, the proper density of the plasma shocked by the forward shock is $\approx\gamma_s n$, its pressure is $p\approx \gamma_s^2 nm_pc^2$, and the energy carried by the plasma shocked by the forward shock is $E_{FS}\approx\gamma_s^2 M_{FS} c^2$ where $M_{FS}=(4\pi/3)n m_p r^3$ is the mass of the plasma shocked by the forward shock. Significant deceleration begins when $E_{FS}\approx E_s$, i.e. when $M_{FS}\approx M_s/\gamma_s$.

Radiation emitted from the shocked plasma at $r_d$ would be observed by a distant observer over $\Delta t\simeq r_d/2\gamma^2_s c^2$. Using this relation with $M_{FS}(r=r_d)\approx M_s/\gamma_s=E_s/(\gamma_s^2c^2)$, we have
\begin{equation}\label{eq:gamma_s}
    \gamma_s = 10^3 \left(\frac{E_{41}}{n_{-1}\Delta t^3_{-4}}\right)^{1/8}, \quad
    r_d= 6\times10^{12}\left(\frac{E_{41}\Delta t_{-4}}{n_{-1}}\right)^{1/4}{\rm cm}.
\end{equation}
Here, the density is $n=10^{-1}n_{-1}{\rm cm^{-3}}$, $E_s=10^{41}E_{41}$~erg and the FRB duration is $\Delta t=10^{-4}\Delta t_{-4}$~s.

Let us consider next the density of the shocked shell plasma at $r=r_d$. The proper density of the un-shocked shell plasma at this point is given by $n_s=M_s/4\pi r_d^2\Delta r_{\rm RF}$, where $\Delta r_{\rm RF}$ is the rest frame thickness of the shell. Since the variation of velocity across the shell is expected to be $\sim c/(2\gamma_s^2)$, the observer frame shell thickness is expected to expand by $\delta\Delta r\approx (r_d/c)c/(2\gamma_s^2)=r_d/(2\gamma^2_s)$ by the time it reaches $r_d$. The initial size of the shell is expected to be $\Delta r\sim R_s$. Assuming that $R_s<r_d/(2\gamma^2_s)=c\Delta t$, we have $\Delta r\approx r_d/(2\gamma^2_s)$ and $\Delta r_{\rm RF}\approx r_d/\gamma_s$ at $r=r_d$. The ratio of the proper density of the un-shocked shell to $n$ is thus $n_s/n\approx \gamma_s M_s/M_{FS}\approx \gamma_s^2$. This implies that the pressure behind the forward shock satisfies $p\approx\gamma_s^2 n m_p c^2\approx n_s m_p c^2$, which in turn implies that the reverse shock becomes mildly relativistic at $r=r_d$, heating the shell protons to $\sim m_p c^2$. This implies that the energy $E$ is roughly equally divided at this stage between the forward and reverse shocked plasmas, and that, as assumed, the reverse shock does not become highly relativistic.

To summarize, at the deceleration radius $r_d$ the characteristic energy per particle and proper density of the shocked shell plasma are
\begin{equation}\label{eq:shell}
    T_s\approx m_pc^2, \quad n_s\approx \gamma_s^2 n= 10^5\left(\frac{E_{41}n^3_{-1}}{\Delta t^3_{-4}}\right)^{1/4}{\rm cm^{-3}},
\end{equation}
and the density of the plasma shocked by the forward shock is $\approx\gamma_s n$.

For a relativistic plasma, the ratio of plasma frequency, $\nu_p^2=ne^2/\pi\gamma_e m_e$, to gyration frequency, $\nu_B=3eB/4\pi\gamma_e m_e c$, is approximately given by $\nu_p/\nu_B\approx(U_e/U_B)^{1/2}=(\epsilon_e/\epsilon_B)^{1/2}$, where $\epsilon_e$ and $\epsilon_B$ are the fractions of the internal energy carried by electrons and magnetic fields. A first principles derivation of the values of $\epsilon_e$ and $\epsilon_B$ obtained in the downstream of collisionless shocks is not yet available. However, observations (in particular of GRBs) as well as numerical simulations (and analytic considerations) imply that the post-shock plasma is not far from equipartition, i.e. that $\epsilon_e$ and $\epsilon_B$ are not far below unity, for mildly to highly relativistic shocks \citep[for a recent review see][]{SironiKeshetLemoine15SSRv}. For characteristic energy of electrons in the shocked shell plasma of $T_s\approx m_p c^2$, corresponding to a Lorentz factor $\gamma_{e,s}\approx m_p/m_e$, we expect $\gamma_{e,s}\gg\sqrt{\nu_p/\nu_B}\approx(\epsilon_e/\epsilon_B)^{1/4}$.

We may now estimate the frequency at which negative reabsorption may be expected \citep[e.g.][]{SW02},
\begin{equation}\label{eq:nuR}
  \nu_{R^*}=\nu_p\sqrt{\frac{\nu_p}{\nu_B}}.
\end{equation}
Assuming that the electrons are near equipartition with the protons, with $\gamma_{e,s}\approx m_p/m_e$, the plasma frequency of the shocked shell plasma is (using $n_s\approx\gamma_s^2n$)
\begin{equation}\label{eq:nup}
    \nu_{p,s}=\sqrt{\frac{n_s e^2}{\pi\gamma_{e,s} m_e}}\approx7\times10^4 \left(\frac{E_{41}n^3_{-1}}{\Delta t^3_{-4}}\right)^{1/8}{\rm Hz},
\end{equation}
and the observed frequency where negative reabsorption may be expected is
\begin{equation}\label{eq:nuRs}
    \gamma_s\nu_{R^*}\approx0.2 \left(\frac{E_{41}n_{-1}}{\Delta t^3_{-4}}\right)^{1/4}\epsilon^{-1/4}_{B,-2}{\rm GHz}.
\end{equation}
Here, the fraction of thermal energy in magnetic fields is $\epsilon_{B}=10^{-2}\epsilon_{B,-2}$.

\subsection{Negative synchrotron self-absorption}
\label{sec:maser}

The synchrotron self-absorption coefficient, taking into account the effects of the plasma, is given for an isotropic electron distribution by \citep{Ginzburg89}
\begin{equation}\label{eq:alpha}
  \alpha^{[p]}_\nu=-\frac{1}{4\pi m_e\nu^2}\int d\gamma_e \gamma_e^2 P^{[p]}_\nu(\gamma_e)\frac{d}{d\gamma_e}\left(\gamma_e^{-2}\frac{dn_e}{d\gamma_e}\right),
\end{equation}
where the synchrotron power per unit frequency emitted by a single electron with Lorentz factor $\gamma_e$ in a polarization $[p]$, $P^{[p]}_\nu(\gamma_e)$, is given by
\begin{equation}\label{eq:Pnu}
  P^{[p]}_{\nu}(\gamma_e)=\frac{2\pi e^2\nu_c}{\sqrt{3}\gamma_e^2c} S^{-1/2}_\nu(\gamma_e)\left[xf^{[p]}(x)\right]_{x=S_\nu^{3/2}\nu/\nu_c}
\end{equation}
with
\begin{equation}\label{eq:fx}
  f^{[\bot,\|]}(x)=\pm K_{2/3}(x)+\int_x^\infty dy\,K_{5/3}(y),
\end{equation}
\begin{equation}\label{eq:nucS}
  \nu_c=\gamma_e^3\nu_B\sin\chi=\gamma_e^2\frac{3eB\sin\chi}{4\pi m_ec},\quad S_\nu(\gamma_e)=1+\left(\frac{\gamma_e\nu_p}{\nu}\right)^2,
\end{equation}
and $\chi$ is the electron pitch angle.

As shown in \citet{SW02}, for electron energy distributions in which $dn_e/d\gamma_e$ rises faster than $\gamma_e^2$ negative reabsorption is obtained at $\nu<\nu_{R^*}\approx(\epsilon_e/\epsilon_B)^{1/4}\nu_p$  (assuming $\gamma_{e}\gg\sqrt{\nu_p/\nu_B}$). The energy distribution of the electrons produced by the mildly relativistic collisionless shock at energies exceeding the characteristic energy, $\gamma_e\gg\gamma_{e,s}$, is expected to be a power-low, typically $dn_e/d\gamma_e\propto \gamma_e^{-2}$. The energy distribution at much lower energy, $\gamma_e\ll\gamma_{e,s}$, is uncertain. In what follows, we estimate the magnitude of the negative reabsorption for a narrow distribution of electrons around $\gamma_{e,s}$ (a high energy extension would not modify the result significantly).

The amplitude of the negative reabsorption coefficient may be estimated in this case by considering a delta function distribution, $dn_e/d\gamma_e=\delta(\gamma_e-\gamma_{e,s})$. In this case, it is straight forward to show that
\begin{equation}\label{eq:alpha_delta}
  \alpha^{[p]}_\nu=\alpha_0F^{[p]}\left[\gamma_{e,s}^2\frac{\nu_B}{\nu_p},\frac{\nu}{\nu_{R^*}}\right],
\end{equation}
with
\begin{equation}\label{eq:alpha0}
  \alpha_0=\frac{\pi}{2\sqrt{3}}\frac{\nu_B}{c}\sqrt{\frac{\nu_B} {\nu_p}},
\end{equation}
and $F$
\begin{equation}\label{eq:F}
  F^{[p]}(g,y)=2y^{-3}\left[f^{[p]}(x)+\left(\frac{1}{2}-\frac{y^{2}}{g}\right)xf'^{[p]}(x)\right]_{x=\tilde{S}y},
\end{equation}
with
\begin{equation}\label{eq:tS}
  \tilde{S}(g,y)=(g^{-1}+y^{-2})^{3/2}/\sin\chi.
\end{equation}

For large values of $g=\gamma^2_{e,s}\nu_B/\nu_p$, $F$ approaches the limit shown in Fig.~\ref{fig}.
\begin{figure}[h]
\epsscale{1} \plotone{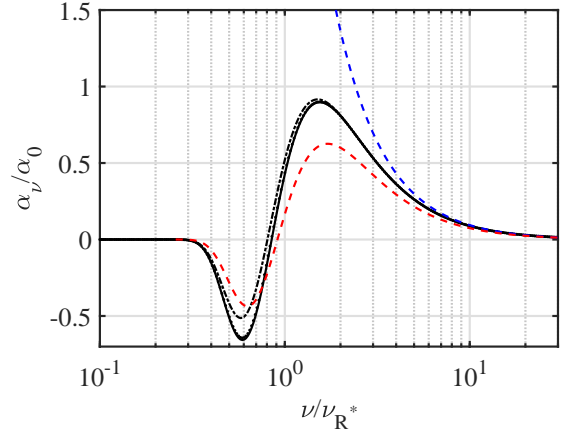}
\caption{The normalized self-absorption coefficient for the $\bot$ polarization (see eqs.~(\ref{eq:alpha_delta}) and~(\ref{eq:alpha0})) and $\sin\chi=1$ as a function of normalized frequency (see Eq.~(\ref{eq:nuR})), for large values of $g=\gamma^2_{e,s}\nu_B/\nu_p$ (dash-dotted, dotted and solid black lines correspond to $g=10,100,1000$ respectively).The dashed red line shows the average over $\chi$ for $g=1000$. The dashed blue line shows the absorption coefficient obtained when plasma effects are neglected.
\label{fig}}
\end{figure}
As expected, $F$ obtains, for large $g$, a minimum of $F\approx -1$ at $\nu/\nu_{R^*}\approx 1$.

We may now obtain the expected exponential factor of amplification due to stimulated emission, $\exp[-\Delta_{\rm RF}\alpha_{\nu}(\nu_{R^*})]$,
\begin{equation}\label{eq:expo}
  -\Delta_{\rm RF}\alpha_{\nu}(\nu_{R^*})\approx -\gamma_s c\Delta t\alpha_0\approx 200\epsilon^{3/4}_{B,-2}(E_{41}n_{-1}\Delta t_{-4})^{1/4}.
\end{equation}
The large amplification factor implies that a significant fraction of the electron energy may be converted to synchrotron maser emission.

Some comments should be made here regarding the validity of the analysis. First, the derivation given above for $\alpha_\nu$ uses the method of Einstein coefficients, leading to Eq.~(\ref{eq:alpha}), instead of solving directly for the zeros of the dielectric tensor. This method is valid as long as the deviations of the dielectric tensor from unity are dominated by the deviations of the refractive index from unity, rather than by the absorption term, i.e. for $|1-n|\gg c\alpha_\nu/\nu$ \citep{Ginzburg89,SW02}. This is satisfied at $\nu\sim \nu_{R^*}$, for which $|1-n|\sim\nu_B/\nu_p$ and $c\alpha_\nu/\nu\sim(\nu_B/\nu_p)^2$. Second, the appropriate polarization modes are the parallel and perpendicular modes of propagation as long as the difference between the refractive indices of the circularly polarized modes introduced by the magnetic field, $\sim \nu_p^2\nu_B/\nu^3$, is small compared to $c\alpha_\nu/\nu$ \citep{Ginzburg89,SW02}. This is satisfied at $\nu\sim \nu_{R^*}$, for which $\nu_p^2\nu_B/\nu^3\sim(\nu_B/\nu_p)^{5/2}$.

Two points should be further considered here. First, as the electrons cool, both the Razin frequency and the amplification factor change. Since $\nu_p\propto\gamma_e^{-1/2}$ and $\nu_B\propto\gamma_e^{-1}$, $\nu_{R*}\propto\gamma_e^{-1/4}$ and $\alpha_{\nu,R}\propto\gamma_e^{-5/4}$. Thus, as the electrons cool the emission slowly shifts to higher frequency and the amplification factor rapidly increases.

Second, let us consider synchrotron self-absorption in the plasma shocked by the forward shock. Noting that the plasma frequency in the plasma shocked by the forward shock is $\nu_{p,s}/\gamma_s$, we find that the optical depth of this plasma is $\tau_\nu\approx10^{-6}(\nu/1~{\rm GHz})^{-5/3}$.

\section{Discussion}
\label{sec:discussion}

\subsection{The persistent source}
\label{summary:persist}

The constraints derived directly from observations on the properties of the persistent source are summarized in the abstract and in \S~\ref{sec:persistent_prop}. We point out below several important conclusions.
\begin{enumerate}
    \item The fact that a self consistent solution, with the upper limit on $R$ derived from source variability consistent with the lower limit derived from the properties of radio emission, and with the upper limit on $t$ derived from the properties of radio emission consistent with the lower limit of 4~yr, is not trivial. It lends support to the assumption that the same source produces both the persistent emission and the FRBs.
    \item The source is nearly resolved, and may be resolved by observing at 10~GHz if its size is close to the upper limit of 0.1~mas. If the source size is close to the lower limit of 0.01~mas, it may be possible to identify a larger refractive scintillation variability at 10~GHz compared to that at lower frequencies (see \S~\ref{sec:persistent_size}).
    \item Radio surveys in nearby galaxies may reveal nearby sources of persistent radio emission similar to the one associated with FRB 121102. The rate of FRBs is not very well known, as the distance out to which they are observable is uncertain. The observed FRB rate, of $\sim10^4/$d, suggests that the birth rate of FRB sources is not far below the supernova rate, implying $\sim0.1 (f/0.1)$ sources per galaxy at an age similar to that of FRB 121102, where $f$ is the ratio of FRB birth rate to the supernova rate.
    \item The flux of the source at higher frequencies, $\gg 10$~GHz, is predicted not to exceed the radio flux.
    \item The fact that $\varepsilon_e\sim m_p c^2$ suggests that the hot radiating plasma was created by the ejection of a mildly relativistic, $M\approx E/c^2\approx 10^{-5}M_\odot$ shell, which propagated into an ambient medium of density $n_c<10^{2.5}R_{17.5}{\rm cm}^{-3}$ or collided with a pre-ejected shell of mass $M_c<10^{-1.5}R_{17.5}^4M_\odot$. The conversion of the fast shell's energy to thermal energy via a collisionless shock would naturally lead to $\varepsilon_e\sim m_p c^2$ and to magnetic field not far below equipartition, as implied by observations.
    \item The plasma could also be produced by the emission of a wind, with $\dot{E}=E/t$ and $\dot{M}\approx\dot{E}/c^2$. The density of the surrounding medium may be decreasing with radius, to $n_c$ at $R$. However, the density should be falling slower than $1/r^2$ in order to confine the hot plasma (an $r^{-\omega}$ density profile provides a speed of sound that decreases outwards for $\omega<3/\gamma_{\rm ad.}=9/5$).
\end{enumerate}

Magnetar-wind models do not lead naturally to the above plasma parameters. The number of radiating electrons determined by Eq.~(\ref{eq:L}) is $N_e>10^{52}$. For the wind luminosity typical of magnetar models, $\sim 10^{41}{\rm erg/s}$ \citep[e.g.][]{Murase16,Dai17Magnetar_no_ejecta,Metzger17Magnetar}, the number of e$^\pm$ pairs expected for multiplicity of $\mu_\pm=10^4$, as commonly inferred for the Crab nebula, is $\sim10^{43.5}t_9$, many orders of magnitude below the required number \citep[$\mu_\pm$ is the ratio between the pair flux carried by the wind and the Goldreich-Julian flux,][]{GoldreichJulian}. Indeed, such models typically predict emission peaking at 10's of keV \citep{Metzger17Magnetar}, or at $\sim10^{15}$~Hz when postulating $\mu_\pm=10^6$ \citep{Murase16}, instead of at 10~GHz as observed. In fact, for strong magnetic fields, as in magnetars, the pair multiplicity is expected to be lower, $\mu_\pm\sim 1$ \citep[e.g.][]{Dai10pairs}, thus exacerbating the discrepancy.

This challenge has been realized by \citet{Beloborodov17Magnetar}, who suggests that the required electron density is produced by the ejection of mass accompanying a giant magnetar flare. Efficient mixing of this mass into the persistent magnetar wind, accompanied by efficient heating of the electrons to near equipartition, may produce the required plasma parameters.

\subsection{The FRBs}
\label{summary:FRB}

We have suggested a mechanism for the generation of FRBs (independent of the persistent source model): Ejection from an underlying compact object, $R_s=10^{6}R_{s,6}$~cm, of highly relativistic shells, with energy $E_s=10^{41}E_{41}$~erg and Lorentz factor $\gamma_s = 10^3 E_{41}^{1/8}R^{-3/8}_{s,6}$, into a surrounding $e-p$ plasma of density $\sim 10^{-1}{\rm cm^{-3}}$. This density is consistent with that inferred for the plasma producing the persistent emission of FRB 121102. Such shell ejections with energy typical for FRB events lead to plasma conditions at the reverse shock driven into the expanding shell, which are appropriate for strong synchrotron maser emission at the GHz range, $\nu_{\rm coh.}\approx 0.5 (E_{41}n_{-1})^{1/4}R^{-3/4}_{s,6}{\rm GHz}$ (see Eq.~(\ref{eq:nuRs})). In this model, the large negative reabsorption amplification factor (see Eq.~(\ref{eq:expo})) implies that a significant fraction of the electron energy may be converted to synchrotron maser emission, observed as an FRB with duration $R_s/c$. Several points are important to emphasize.
\begin{enumerate}
    \item Negative reabsorption is obtained for a narrow range of frequencies, see Fig.~\ref{fig}.
    \item As the radiating electrons cool, we expect a slow rise in the emission frequency.
    \item Since the FRB carries a significant fraction of the energy, stronger emission at other wavelengths is not expected. It is straight forward to show that synchrotron emission from the forward shock driven by the shell is expected to produce a burst of $\sim 10$~MeV photons carrying energy, which is smaller than that of the FRB and unlikely to be detectable. This is in contrast with the widely discussed magnetar flare scenario (see below), in which the FRB is accompanied by a high energy photon burst carrying $\sim10^8$ times more energy than the FRB.
    \item Negative reabsorption requires an electron energy distributions $dn_e/d\gamma_e$ which rises faster than $\gamma_e^2$ below the characteristic electron energy ($\gamma_{e,s}\sim m_p c^2$) in the plasma shocked by the mildly relativistic reverse shock. The energy distribution of the electrons produced by the mildly relativistic collisionless shock at energies exceeding the characteristic energy is expected to be a power-low, typically $dn_e/d\gamma_e\propto \gamma_e^{-2}$. The energy distribution at much lower energy, $\gamma_e\ll\gamma_{e,s}$, is uncertain. We have estimated the magnitude of the negative reabsorption for a narrow distribution of electrons around $\gamma_{e,s}$ (a high energy extension would not modify the result significantly). The viability of the model presented here for the FRBs depends on the validity of the assumption that $dn_e/d\gamma_e$ rises faster than $\gamma_e^2$ at low energy, and on the absence of plasma instabilities that may grow faster than the maser instability and thus quench the maser emission discussed here \citep[note that negative reabsorption may be present for an isotropic plasma distribution, which is stable against the electrostatic "two stream" instability discussed in the FRB context in][]{Katz16two_stream}. This requires further investigation.
    \item The frequency at which coherent emission is expected, and the amplification factor, are determined by the energy and duration of the FRB event. The fact that for values typical for FRB events the coherent emission frequency is predicted to be in the GHz range, with strong amplification, lends some support to this model.
    \item Since the ejected shells are required to be highly relativistic, $\gamma_s\gg1$, the ejection of shells which are not spherical, but rather conical sections of a sphere with opening angle $\theta_s\gg 1/\gamma_s$, would lead to similar observed burst properties (as long as our line of sight to the source falls within the cone): A conical shell with $\theta_s\gg 1/\gamma_s$ behaves as if it were part of a spherical shell (until significant deceleration sets in). Thus, $E_s$ should be regarded as the "spherically equivalent energy", while the true energy emitted is $\sim\theta_s^2 E_s$.
\end{enumerate}

It is widely believed that a coherent emission mechanism that may produce the FRBs may operate within a "magnetar wind nebula", i.e. within an e$^\pm$ plasma generated by a highly relativistic magnetized wind emitted from a fast rotating and highly magnetized, $B\sim10^{14}$~G, neutron star, and confined by a supernova ejecta \citep[see][for alternative scenarios]{Geng15steroid,Mottez15pulsar-planet,Gu16NS-WD,Dai16asteroid,Romero16,Zhang17comb}. However, the physics of such a mechanism is not yet understood \cite[see][for discussion]{Katz16Rev}. A specific model, in which FRBs accompany  "giant magnetar flares", was suggested by \citet{Lyubarsky14maser}. In this model, the flare drives a collisionless shock into the pair-plasma, and coherent synchrotron emission is produced at the gyration frequency of the electrons due to an anisotropic particle distribution at the shock front (it is worth noting in this context that synchrotron maser radio emission due to anisotropic relativistic electron distribution has been suggested to operate in astrophysical plasmas by \citet{Sazonov73}, and reintroduced more recently in the context of FRBs by \citet{Ghisellini17}).

A significant challenge faced by this model is related to the fact that the flare energy, $\sim10^{48}$~erg, is $\sim10^9$ times larger than the FRB energy (\citet{Lyubarsky14maser,Metzger17Magnetar}; A variant of this model, where the ratio between the flare energy and the FRB energy is smaller, $\sim10^{5}$, has been proposed by \citet{Beloborodov17Magnetar}). It is thus challenging to explain within this model the FRB energy without a complete detailed understanding of the physics, including the physics of collisionless shocks, which is not yet in hand. The model is further challenged by FRB 121102 observations, since for model parameters typically used to construct models for this FRB, $B\sim10^{14}$~G and rotation frequency $\Omega\sim10^2{\rm s^{-1}}$, maser emission is predicted at $\nu<1$~MHz.

\subsection{The underlying compact object}
\label{sec:source}

The inferred energy and mass associated with the source are somewhat low compared to those of typical supernova ejecta (A similar conclusion has been reached in recent analyses of FRB 121102 \citep{Murase16,Piro16,Kashiyama17,Beloborodov17Magnetar,Dai17Magnetar_no_ejecta,Katz17dark}, suggesting that the source is a magnetar-wind nebula confined by a (very) low mass supernova ejecta). This may suggest some type of a "weak stellar explosion", where a neutron star is formed with relatively low mass and energy ejection (and hence possibly not associated with a typical supernova remnant). Plausible candidates may be accretion induced collapses of WDs or WD mergers. However, current observations do not allow one to rule out $M_c\sim1M_\odot$, or the ejection of larger mass well before the ejection of the confining shell.

The formation of the neutron star may plausibly be accompanied by the ejection of a mildly relativistic shell of energy $\sim10^{49}$~erg, thus generating the plasma producing the persistent emission. The ejection of the highly relativistic shells, that may give rise to the FRBs, requires a separate explanation. Note, that the time averaged luminosity of the FRBs is not high, $\sim10^{34}\theta_s^2{\rm erg/s}$, implying that the total energy carried by the highly relativistic shells over the persistent source life time is $\lesssim 10^{44}\theta_s^2{\rm erg}$.

\acknowledgements  I would like to thank Eran Ofek, Boaz katz, Doron Kushnir, Scott Tremaine, Tejaswi Venumadhav Nerella, and the anonymous referee for useful discussions. This research was partially supported by an ISF I-Core grant and an IMOS grant.

\bibliographystyle{hapj}

\end{document}